\documentclass[runningheads,a4paper]{llncs}

\usepackage{amssymb}
\usepackage{graphicx}
\usepackage{amsmath}
\usepackage{tabularx}

\usepackage{bibnames}

\usepackage{url}
\urldef{\mailsa}\path|{amanmeet_garg,dla121,kpopuri,mfbeg}@sfu.ca|   
\newcommand{\keywords}[1]{\par\addvspace\baselineskip
\noindent\keywordname\enspace\ignorespaces#1}

\begin{document}

\mainmatter  % start of an individual contribution

\title{Cortical Geometry Network and Topology Markers for Parkinson's Disease}

\titlerunning{Geometry Network Topology Marker}

%\author{Alfred Hofmann%
% \thanks{Please note that the LNCS Editorial assumes that all authors have used
% the western naming convention, with given names preceding surnames. This determines
% the structure of the names in the running heads and the author index.}%
% \and Ursula Barth\and Ingrid Haas\and Frank Holzwarth\and\\
% Anna Kramer\and Leonie Kunz\and Christine Rei\ss\and\\
% Nicole Sator\and Erika Siebert-Cole\and Peter Stra\ss er}
%
\author{Amanmeet Garg \and Donghuan Lu \and Karteek Popuri \and Mirza Faisal Beg}

\authorrunning{Garg et al.}

%\institute{Anonymous
\institute{School of Engineering Science,
Simon Fraser University, Canada\\
\mailsa\\
%\url{http://www.springer.com/lncs}
}

\toctitle{Lecture Notes in Computer Science}
\tocauthor{Authors' Instructions}
\maketitle

\begin{abstract}
Neurodegeneration affects cortical gray matter leading to loss of cortical mantle volume. As a result of such volume loss, the geometrical arrangement of the regions on the cortical surface is expected to be altered in comparison to healthy brains. Here we present a novel method to study the alterations in brain cortical surface geometry in Parkinson's disease (PD) subjects with a \emph{Geometry Networks (GN)} framework. The local geometrical arrangement of the cortical surface is captured as the 3D coordinates of the centroids of anatomically defined parcels on the surface. The inter-regional distance between cortical patches is the signal of interest and is captured as a geometry network. We study its topology by computing the dimensionality of simplicial complexes induced on a filtration of binary undirected networks for each geometry network. In a permutation statistics test, a statistically significant ($p<0.05$) difference was observed in the homology features between PD and healthy control groups highlighting its potential to differentiate between the groups and their potential utility in disease diagnosis.

\end{abstract}

% -----------------------------------------------------------------------------------------------
% -----------------------------------  INTRODUCTION  --------------------------------------------
% -----------------------------------------------------------------------------------------------

\section{Introduction}
Computational techniques of network analysis applied to neuroimaging data have helped to characterize the changes associated with Parkinson's Disease (PD) in the structural \cite{Zheng2014}, functional \cite{OldeDubbelink2014a} and morphological networks \cite{Pereira2015a}. Brain shrinkage (reduction in volume) and cortical gray matter loss have been observed in PD patients \cite{Melzer2012}  suggestive of changes in the geometric arrangement of brain regions. Yet, the relative geometric arrangement of the brain structures is often overlooked as a potential marker for neurodegenerative diseases. In this work we present a novel framework to quantify the geometric arrangement of brain regions to study its alterations in the presence of a neurodegenerative disease such as PD.

Current network analysis of brain data is based on the simplifying assumption of a dyadic (pairwise) interaction between nodes (regions of brain) of the network \cite{Rubinov2010}. Such an assumption potentially limits the underlying structural and functional interactions captured by the brain graph models. The general framework of topology data analysis enables the study of polyadic (one-to-many) interactions via application of principles rooted in the theory of simplicial complexes and persistent homology \cite{Carlsson2009,Mischaikow2013}. A similar research work studied brain changes via topology features \cite{Lee2014}, however, their work focused on group level graphs with subjects as nodes. In this work we present a framework to quantify and characterize the geometrical arrangement of the cortical surface patches via a morphological network for an individual. Further, we compute the algebraic topology features to capture the multiway interactions between networks nodes.

Our geometry network (GN) framework encodes the geometric arrangement of cortical anatomical patches with an inter-region Euclidean distance in the 3D space. Further, we induce a filtration of Vietoris-Rips complexes on the inter-region distance matrices to obtain the persistence information and the counts of the k-simplices forming the complex (explained in section \ref{subsec:PersHomo}). These features are then tested for their ability to differentiate between groups. We demonstrate the applicability of this framework to T1 structural MRI data for a population of PD patients and healthy control subjects, however, it is general in nature and can be extended to other modalities and neurodegenerative diseases. The key contributions of our work are two fold, 1) a study of the geometrical arrangement of cortical brain regions as affected by Parkinson's disease and 2) a novel framework to characterize the polyadic interactions between brain regions to obtain discriminative features.

% -----------------------------------------------------------------------------------------------
% ---------------------------------------  METHODS  ---------------------------------------------
% -----------------------------------------------------------------------------------------------
\section{Methods}
\label{Sec:methods}

Our framework is based on the premise that the inter-regional geometrical arrangement in the brain is altered by the disease related neuro-degeneration. The method we present here focuses on the alterations in cortical surface geometry approximated as a set of anatomically defined patches, whose centroids are representative of their geometric localization in the 3D space. Further, the topological analysis takes the inter-regional (node-to-node) Euclidean distance information and induces filtration of simplicial complexes to obtain features of persistent homology for the brain point cloud. Below we describe our method in further detail along with a workflow outline (Figure \ref{fig:BlockDiag}).

% -----------------------------------------------------------------------------------------------
\subsection{Geometry Networks}
\label{subsec:GeomNW}
The central aim of our work is to quantify the geometrical arrangement of the cortical structures in the brain. Thus, their localization in the 3D space is the signal of interest and is captured in the polyadic (many-to-many) interaction via the Euclidean distance between the patch centroids. Such $n\times n$ inter-patch distance matrices for $n$ nodes enable us to study the graph theoretic and topology properties of the geometrical arrangement and compare them across groups.

% -----------------------------------------------------------------------------------------------
\paragraph*{Network Filtration:}
In network analysis, the threshold to obtain a binary graph from a weighted graph is a parameter often selected based on the suitability to the application at hand, thus, limiting its generalizability. To circumvent this issue, we construct a network filtration with a monotonically increasing set of threshold values to obtain a set of binary undirected networks with different levels of network sparsity. This approach has the advantage of providing a complete set of network topology characteristics from a fully disconnected network to a fully connected network.

A weighted undirected brain graph $G$ is thresholded at a value $\varepsilon_k$ to yield a binary undirected graph $G_k$. Upon changing the threshold $\varepsilon_1 < \varepsilon_2 < \cdots < \varepsilon_k< \cdots < \varepsilon_n$ we get a hierarchical sequence of $n$ binary undirected graphs $G_1 \subseteq G_2 \subseteq \cdots \subseteq G_k \subseteq \cdots \subseteq G_n$ termed as a \emph{Network Filtration}. A graph originates from a distance matrix $D$ where, each entry $d_{ij}$ is the connection strength (Euclidean distance for geometry networks) between the nodes $i$ and $j$. Each $D$ is converted into an adjacency matrix $A_k$ where, $\{a_{ij} = 1 | d_{ij} < \varepsilon_k, 0\,\, otherwise\}$ for a chosen threshold $\varepsilon_k$ giving the graph $G_k$. We compute the features of nodal degree, betweenness centrality, clustering coefficient, local efficiency and subgraph centrality for each network in the filtration. For a detailed mathematical description of these features, the reader is guided to the seminal work by Rubinov and Sporns \cite{Rubinov2010}. 

The inter-patch distance for each subject varies depending on the brain size rendering the graph filtration generated based on the raw distance values influenced by the scale of the overall brain size in addition to the relative inter-regional distances whose alterations with disease are of interest. Thus, in order to overcome such scale-related differences, we normalize the values of the inter-patch distance for each individual to the range of $[0,1]$ prior to the generation of a graph filtration. This enables a comparison of the network and topology features across subjects and groups by potentially reducing the affect of scale variation of the brain size. 

% -----------------------------------------------------------------------------------------------
\subsection{Persistent Homology}
\label{subsec:PersHomo}
The central idea behind the theory of persistent homology (PH) is to build a sequence of nested subsets on a space of simplicial complexes, studied at different resolutions. For our work, the Vietoris-Rips ($VR$) complex completely defined by the underlying 1-skeleton is induced on a symmetric distance matrix of pairwise distances between points in a point cloud.

A VR complex is defined on a metric space $M$ for a specific distance value $\delta$ by forming a $k$-simplex for every finite set of $k+1$ points that has diameter at most $\delta$. For a set of $p$ nodes in the point cloud, the VR complex has at most $(p-1)$ simplices, enabling the geometry networks to obtain higher dimensional interactions limited in binary networks to 1-dimensional simplices (edges). We characterize the topology of the point cloud at a given threshold $\varepsilon_k$ by a Betti number feature where the $m^{th}$ Betti number ($\beta_m$) is the count of $(m+1)$ simplices in a point cloud \cite{Carlsson2009}. Monotonically increasing values of the scale parameter $\varepsilon_k$ lead to a $VR$ filtration where $VR_{\varepsilon_1} \subseteq VR_{\varepsilon_2} \cdots \subseteq VR_{\varepsilon_k} \cdots \subseteq VR_{\varepsilon_n}$. Akin to the network filtration obtained above (section \ref{subsec:GeomNW}) we obtain the PH features for the VR filtration from the geometry network distance matrices. We compute the Betti numbers for the connected components (Betti-0, $\beta_0$), tunnels (Betti-1, $\beta_1$), cavities (Betti-2, $\beta_2$) and the 4-dimensional simplices (Betti-3, $\beta_3$). As a result, for each threshold value in the filtration we obtain a Betti number (presented in the Figure \ref{fig:PH_Features_Distribution}).

We input the distance matrices for the geometry networks in the package \textit{Perseus} with the parameter (\textit{distmat}) to compute the PH of the Vietoris-Rips complex for each brain point cloud from the inter-patch distance matrices \cite{Mischaikow2013}. The algorithmic basis of the package uses discrete Morse theory to reduce the size of the simplicial complexes to efficiently compute the PH features. We obtain the Betti numbers ($\beta_i , i = 0,1,2,3$) for the filtration for each subject with threshold $\varepsilon_k = 0,0.01,0.02, \cdots,1$ resulting in a total of 100 values.

In order to interpret the Betti numbers in relation to the geometry networks of the brain it is important to understand the underlying information represented by these numbers. The Betti-0 number represents the number of connected components or nodes. The separateness of a point cloud with high Betti-0 number values suggests a higher separation. The number of tunnels (Betti-1) suggest the density of geometry information being shared in local neighborhood of the point clouds. The number of voids (Betti-2) indicates the degree of segregation of the connections, with higher values in the disease group suggesting a more local or segregated co-variation of the geometry. Similarly, the Betti-3 suggests interaction of 4 or more nodes at a given threshold.

\subsection{Image processing}
\label{subsec:image_processing}

\paragraph*{Cortical segmentation: }
The outer surface boundary for the cortical gray matter (pial surface) was obtained from the FreeSurfer 5.1 method \cite{Fischl1999}. The surfaces were quality controlled for segmentation accuracy and, incorrect segmentations were corrected manually where possible or removed from further analysis. The final cortical surface model for each subject is represented as a triangulated mesh with vertices and faces.

\paragraph*{Adaptive Cortical Parcellation: }
As the first step in our method we normalize the cortical surface of the individual target subjects to an average template to obtain vertex wise correspondence in the cortical mesh across the database. %The average template for this work was obtained via an iterative registration and fusion of T1 MRI images from 98 out-of-study control subjects. 
The cortical surface thus obtained was further subdivided into patches constrained to be within the anatomical parcellations obtained from the Freesurfer method \cite{Fischl1999}. A k-means clustering of the vertex coordinates was performed to obtain a patch-based representation of the cortical surface. An additional constraint on the minimum number of vertices per patch (\textit{mvcpp}) was included to avoid extremely small or large patches \cite{Raamana2013}. The cortical patch-wise parcellation from the template surface was transferred to individual target subjects in their respective spaces with spherical daemons registration \cite{Fischl1999}. The coordinates of the centroid for each patch were obtained as the average of the coordinates of the vertices within each patch.

It is important to note that \textit{mvcpp} is a parameter of choice and directly influences the number of patches on a cortical surface. A higher value of \textit{mvcpp} leads to larger patches in the k-means clustering resulting in fewer total number of patches $N$. This results in a smaller-sized distance matrix per brain $N \times N$. A very large value of mvcpp would give smaller distance matrices, however could lead to a potential loss of information. On the contrary a very small mvcpp would lead to more patches giving larger distance matrices which may become memory and time inefficient. 

% -----------------------------------------------------------------------------------------------
\paragraph*{Skull Normalization: }
In order to study the the geometrical arrangement of the brain regions, it is vital to reduce the variability across subjects due to head size. The brain development into the adult healthy brain follows the space restriction enforced by the bony cranial vault and hence we assume that the brain geometrical shape is contoured along the shape of the cranial vault. The first step of normalization uses a 12 degree of freedom affine transformation between the `cranial vault shapes' as it offers an anatomically-meaningful way to normalize the geometry based on the estimated contour of each brain in its adult disease-free state. This affine-based normalization can be interpreted as an extension of intra-cranial vault volume (ICV)-based scaling that is commonly used to normalize volumes of brain structures before comparison across the database. The affine transformations estimate potentially different scaling in the $x,y$ and $z$ directions based on registering the cranial-vault shapes which could be important when the goal is to study the relative geometrical arrangement of brain regions, and not just the volumes of brain regions. The vertices on the surface of the cortical gray matter (pial surface) were then transformed with this affine transformation to reduce the effects of head size-related variability across subjects.

% --------------------------------------------------------------------------------------------

\subsection{Statistical Analysis}
\label{subsec:StatAnalysis}
The classical network features (1 value per node, 140 nodes x 100 thresholds = 14000 dimensions) and Betti number features (100 dimension) live in a high dimensional space relative to the sample size in our dataset (339 PD, 150 CN). In order to minimize the effects of the curse-of-dimensionality (small sample size, high dimensionality), we mapped the features into a reduced dimensionality space created by the principal components of the PCA decomposition of these features. Data for a feature (Local efficiency or Betti-3 etc.) from all subjects in the two groups were mapped into its reduced space of $m$ dimensions. We retained $m$ PCs sufficient to account for 95\% of the variability in the original features across the dataset. Consequently, each feature obtained a different number of PCs (Table \ref{Tab:Stats_Features}).

The ability of the network and \emph{PH} features (mapped into this new dimension space (principal component loadings)) to differentiate between the group of patients and healthy individuals was tested via a Fisher's exact test with random group assignment. The test was repeated for 20,000 random permutations to obtain an empirical estimate of the data distribution to compare the observed group level difference of the features. The component scores for each subject for each individual or combination of features were used in a Fisher's exact test to check for group level difference.

For each permutation we compute the Hotelling's $T^2$ statistic, which takes the form 
\begin{equation}
T^2 = \frac{N_{Grp1} N_{Grp2}}{N_{Grp1} + N_{Grp2}} \times \left(M^{Grp1} - M^{Grp2}  \right)^T \hat{\Sigma}^{-1} \left(M^{Grp1} - M^{Grp2}  \right)
\end{equation}
where, $m$ PCs are selected and each group has $N_{Grp}$ subjects with a mean value of $M^{Grp}$. This gives rise to the empirical distribution $\hat{F}(\cdot)$ as per
\begin{equation}
\hat{F}_{m,N_{Grp1}+N_{Grp2}-m-1} = \frac{N_{Grp1}+N_{Grp2}-m-1}{(N_{Grp1}+N_{Grp2}-2)m} T^2
\end{equation}
where, the null hypothesis of two groups having equal distribution is rejected when $p = \int \limits_{T^2}^{\infty} \hat{F}(f) \mathrm{d}f$ falls below the predefined significance level ($p<0.05$).

% ******************************************************************************

\subsection{The framework}
\label{subsec:framework}
\begin{figure}[h!]
\centering
\includegraphics[width=\textwidth]{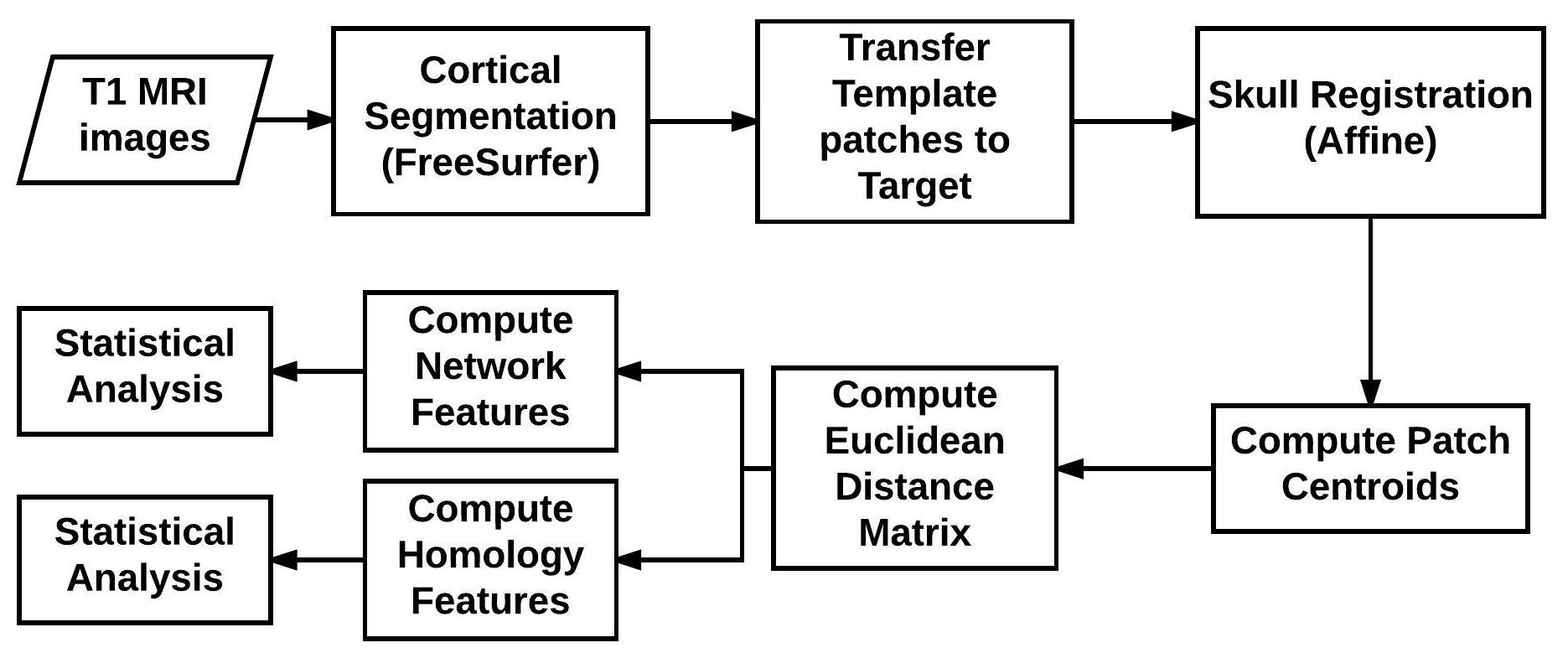}
\caption{Block Diagram workflow of the proposed Geometry Network (GN) framework.}
\label{fig:BlockDiag}
\end{figure}

In order to compute the classical graph theoretic and persistent homology features from the T1 MRI data, we followed a sequence of steps as outlined in the figure \ref{fig:BlockDiag}. Triangulated cortical surface meshes (section \ref{subsec:image_processing}) were normalized to obtain vertex wise correspondence between subjects across the groups (section \ref{subsec:image_processing}). Further, these surfaces were normalized for head size differences between subjects via normalization of cranial vault shape by affine registration to a common template (section \ref{subsec:image_processing}). These surfaces were then used to compute the geometry networks, classical network features (section \ref{subsec:GeomNW}) and persistent homology features (section \ref{subsec:PersHomo}) followed by test for group level difference between the features in a Fisher's exact test(section \ref{subsec:StatAnalysis}).

\subsection{Materials}
Imaging data for this work was obtained from the publicly available database provided under the Parkinson's Progressive Markers Initiative (\emph{PPMI}). Detailed protocol for image acquisition and quality control for the study is available at the website \emph{www.ppmi-info.org}. The two groups with De Novo PD patients (n = 339, age = 61.28$\pm$9.84, 209M/130F) and healthy controls (CN) (n = 150, age = 59.60$\pm$11.33, 93M/57F) were selected and analysed through the method as described in section \ref{subsec:framework}.

% -----------------------------------------------------------------------------------------------
% ---------------------------------------  RESULTS  ---------------------------------------------
% -----------------------------------------------------------------------------------------------

\section{Results}
\label{sec:Results}
A template parcellation into patches with \textit{mvcpp} = 2000 was transferred to individual subject cortical surfaces. This resulted in $70$ patches per hemisphere for a total of $N=140$ patches per brain. Computing the centroid coordinates of each patch and the Euclidean distance between the patch centroids resulted in 489 (339 PD, 150 CN) distance matrices of size $N \times N = 140 \times 140$.

\begin{figure}[htb!]
\centering
\includegraphics[width=\textwidth]{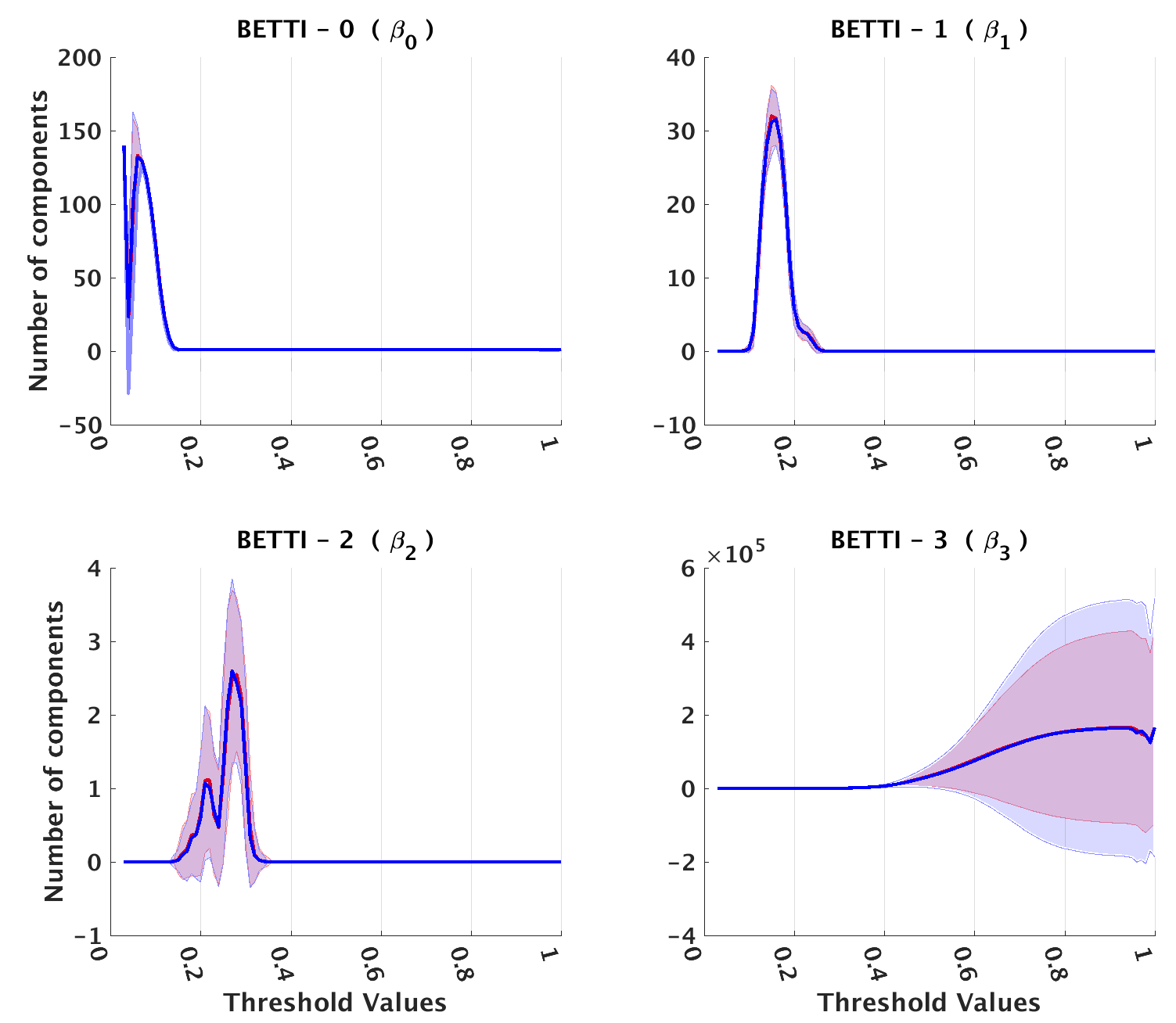}
\caption{Geometry network persistent homology features for the Parkinson's disease (red) and healthy control group (blue). The mean value of the betti numbers are drawn with solid lines with standard deviation from the mean shown with shaded areas.}
\label{fig:PH_Features_Distribution}
\end{figure}

The PH features from the VR filtration from each brain showed a variation over filtration threshold for the betti numbers and across the groups (Figure \ref{fig:PH_Features_Distribution}). The mean values show overlap between the two groups however the variation in the groups shows visual difference (shaded areas in Figure \ref{fig:PH_Features_Distribution}). It is important to note that the distribution of the betti numbers aligns with our interpretation as with increasing threshold in the filtration, we see a very early demise of Betti-0, followed by Betti-1 and Betti-2. Additionally, the Betti-3 numbers show a late appearance and continue to increase in value to the highest threshold value ($\varepsilon_k = 1$). %Likewise, the classical network features (Figure \ref{fig:NWFeat}, mean z-scores in the PD group) show a small value of the mean z-score for a large subset of threshold values suggesting a relatively small difference between the two groups.

\begin{figure}[h!]
\centering
  \begin{tabular}[]{c}
\includegraphics[width=0.85\textwidth]{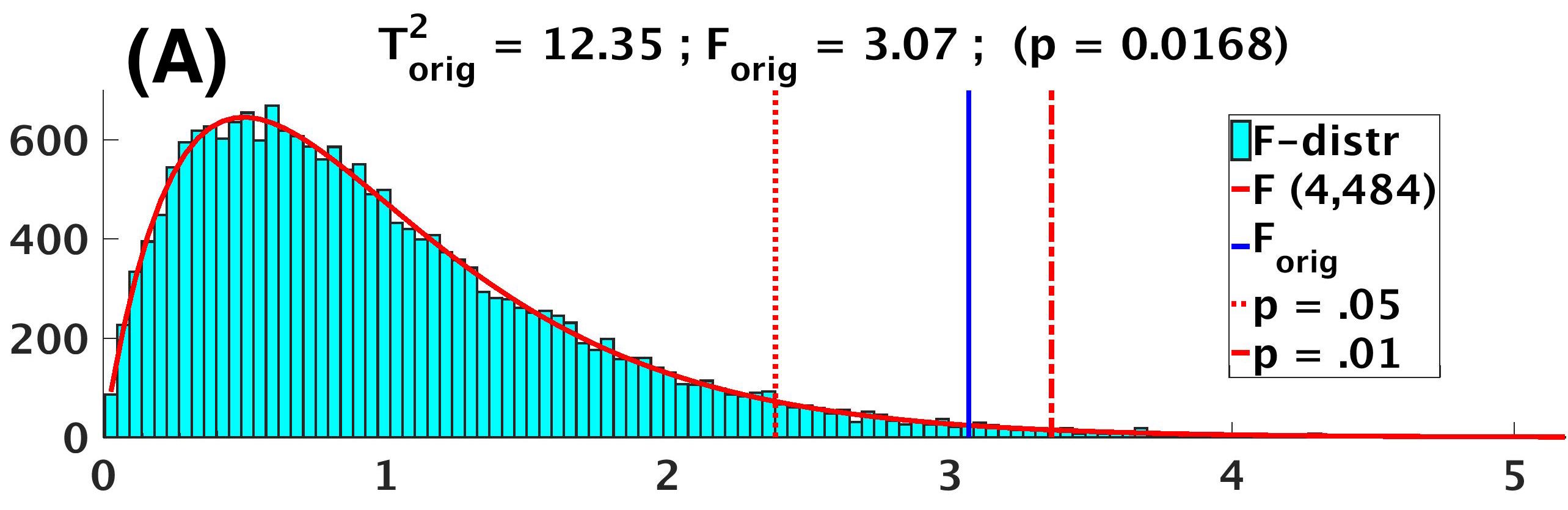} \\ 
\includegraphics[width=0.85\textwidth]{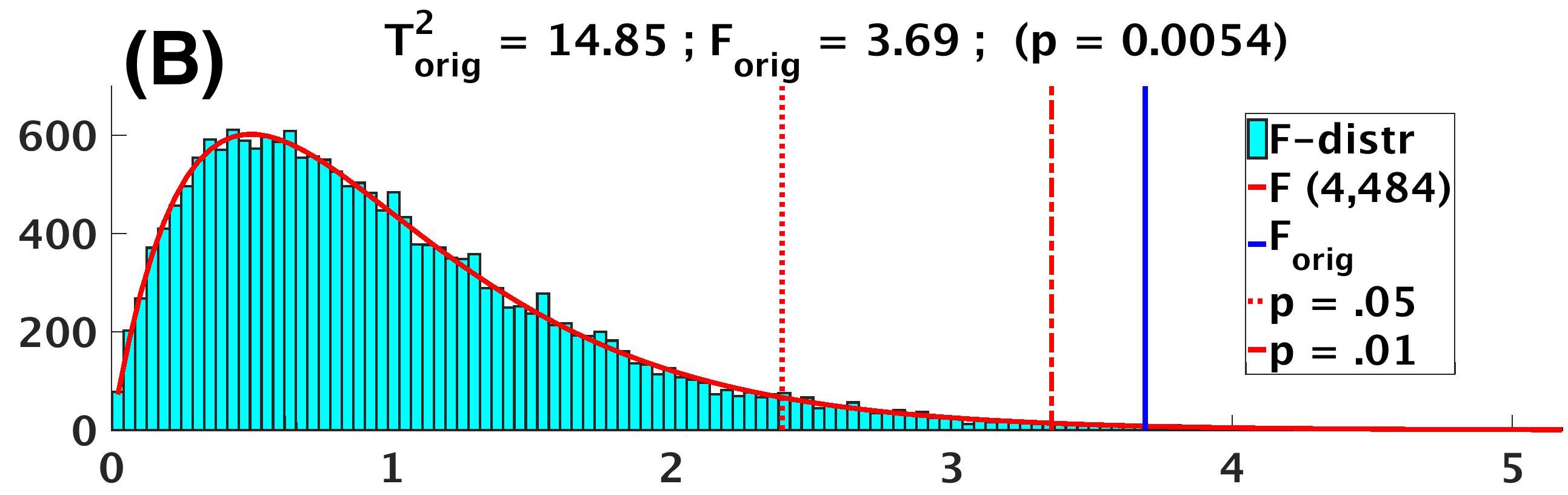}  \\
\end{tabular}
\caption{Results for the Fisher's exact test for group level difference between the Parkinson's patient and healthy control groups in the A) Betti-1 ($\beta_1$) and B) Betti-3 ($\beta_3$) features.} 
\label{fig:PH} 
\end{figure}

The network features in the reduced PCA dimension were unable to differentiate (in a statistical sense) between the geometry networks in the PD and CN groups (Table \ref{Tab:Stats_Features}). The PCA decomposition of the $\beta_0$ number selected the first PC suggesting a strong similarity across the groups, reiterated by the statistical test ($p>0.05$, Table \ref{Tab:Stats_Features}). The Betti numbers $\beta_1$ and $\beta_3$ showed statistically significant difference between the two groups in the permutation experiment ($p>0.05$, Figure \ref{fig:PH}, Table \ref{Tab:Stats_Features}). Combining the $\beta_1$ and $\beta_3$ numbers showed improvement in the discrimination performance.

% ---------------- Figure -----------------------------------------------------------

% \begin{figure}[htb!]
%   \begin{tabular}[]{c}
%        \vspace{-1em} \\
% \includegraphics[width=\textwidth]{NWFeat_GeomNW_V1.jpg} \\ 
% \end{tabular}
% \caption{Geometry network classical network features in the Parkinson's disease group. The color patches represent the mean value of the normalized z-scores in the PD group. The nodes vary along the y-axis and the thresholds vary along the x-axis and color patch in each cell represents the mean z-score value according to the colormap.} 
% \label{fig:NWFeat} 
% \end{figure}

% --------------- Table ------------------------------------------------------------
\begin{table}[h]
\centering
\textbf{\refstepcounter{table}\label{Tab:Stats_Features} Table \arabic{table}.}{ Group difference statistics (PD vs. CN) results for the classical network and persistent homology features for the geometry networks ($p<0.05$).}
\begin{tabular*}{\textwidth}{c @{\extracolsep{\fill}} cccccc}
\hline
	Feature & Num. of PCs & p-value & $T^2$ & F \\ \hline
    Subgraph centrality		&	1	&	0.416	&	0.67	&	0.67	\\
    Local Efficiency		&	109	&	0.182	&	159.06	&	1.14	\\
    Clustering Coefficient	&	139	&	0.341	&	203.85	&	1.05	\\
    Betweenness Centrality	&	49	&	0.481	&	54.08	&	1.00	\\
    Nodal Degree			&	3	&	0.242	&	4.16	&	1.38	\\ 
    Betti-0					&	1	&	0.189	&	1.77	&	1.77	\\
    Betti-1					&	4	&	0.017*	&	12.34	&	3.07	\\
    Betti-2					&	8	&	0.849	&	4.15	&	0.51	\\
    Betti-3					&	4	&	0.005*	&	14.85	&	2.87	\\
    Betti-1 \& Betti-3		&	5	&	0.005*	&	14.82	&	3.03	\\     \hline
\end{tabular*}
\end{table} 

% -------------- Figure -------------------------------------------------------------

% -----------------------------------------------------------------------------------------------
% ---------------------------------------  DISCUSSION -------------------------------------------
% -----------------------------------------------------------------------------------------------
\section{Discussion and Future Work}
The current work presents a novel \emph{Geometry Networks} framework as a non-invasive imaging marker for neurodegenerative disorders and supporting statistical evidence of its ability to differentiate between patient (Parkinson's disease) and healthy groups. These networks capture the 3D geometrical arrangement of the anatomically defined cortical (gray matter) regions and the change in this geometry with disease. Graph theoretic and persistent homology features of these data were obtained for corresponding representations as they encapsulate the dyadic and polyadic interactions between the nodes of the network and the points in the 3D point cloud respectively. Statistical evidence presented in our work (section \ref{sec:Results}, Table \ref{Tab:Stats_Features}) highlights, the ability of the homology features to capture the subtle changes in the geometry of the human cortex with the affects of disease. Further, the evidence suggests that the geometrical arrangement of the cortical regions is influenced by PD related neurodegeneration.

It is important to note that the network features (section \ref{subsec:GeomNW}) were unable to show a statistically significant difference between the groups (Table \ref{Tab:Stats_Features}) indicating perhaps the limitation of the simplifying assumption of a dyadic interactions between the nodes of the networks. On the contrary, the persistent homology features of Betti-1 ($\beta_1$) and Betti-3 ($\beta_3$) numbers showed statistically significant difference ($p<0.05$, section \ref{sec:Results}, Table \ref{Tab:Stats_Features}) between the groups (PD vs. CN) in a rigorous Monte Carlo randomization based Fisher's exact test (section \ref{subsec:StatAnalysis}). This suggests that the etiology of Parkinson's disease affects the multi-way interaction between points in the 3D point cloud where as the dyadic interactions alone show small variation from the healthy control population.

% These results suggest that features derived from the geometrical arrangement of the cortical regions of the brain may reveal signatures of neurodegeneration in the brain. Additionally, in contrast to previous work \cite{Lee2014}, our framework generates the persistence homology features from the filtrations of Vietoris-Rips complexes induced on brains for each individual subject. Thus, on a group level we can compare subjects in a randomization based statistical test to check for true group level difference. This further enables the potential use of the features as a computational anatomy biomarker for each subject.

In this work we selected a \textit{mvcpp} = 2000 for adaptive parcellation of the cortical surface. This parameter choice was guided by the computational limitation of time and memory requirements for computing homology features for larger distance matrices. Further research into algorithmic development is needed to enable computation of homology of larger matrices thus, enabling more precise geometrical assessment of the cortical surface.

As geometry and size of the brain change with age and development, the framework presented in this work provides with the opportunity to study the geometric arrangement in different age groups, its change with age and potentially detrimental affects of disease. The results of this work lead to the future work of extending the framework to consider a combination of cortical anatomy with subcortical and white matter anatomical regions to build a whole brain geometrical network. Considering cortical-surfaces derived from longitudinal within-subject images, inter-patch distances across these time-indexed patches could provide signatures of geometrical changes over time.  Additionally, development of machine learning models suitable of processing these features may lead to development of insights that could form part of future computerized diagnostic systems.

\subsubsection*{Acknowledgment}
Data used in the preparation of this article were obtained from the Parkinson’s Progression Markers Initiative (PPMI) database \\ (http://www.ppmi-info.org/data). For up-to-date information on the study, visit www.ppmi-info.org. The authors acknowledge and thank the Natural Sciences and Engineering Research Council (NSERC), Michael Smith Foundation for Health Research (MSFHR), Canadian Institute of Health Research (CIHR), Brain Canada and MITACS Canada for their generous funding support.

\bibliographystyle{splncs03}

\bibliography{Bibliography.bib}

\end{document}